\documentclass[preprint2]{aastex}
\usepackage{spr-astr-addons}
\usepackage{graphicx}
\begin{document}
%Insert your title here
\title{The effect of magnetic fields on the formation of circumstellar discs around young stars}

\shorttitle{MHD and circumstellar disc formation}        % if too long for running head
\shortauthors{Price \& Bate}

\author{Daniel J. Price and Matthew R. Bate}
\affil{School of Physics, University of Exeter, Stocker Rd, Exeter EX4 4QL, UK}

%\email{aastex-help@aas.org}

%% Notice that the first 2 authors have an alternate affiliation, which
%% is identified by the \altaffilmark after each name.  Specify alternate


\begin{abstract}
We present first results of our simulations of magnetic fields in the formation of single and binary stars using a recently developed method for incorporating Magnetohydrodynamics (MHD) into the Smoothed Particle Hydrodynamics (SPH) method. An overview of the method is presented before discussing the effect of magnetic fields on the formation of circumstellar discs around young stars. We find that the presence of magnetic fields during the disc formation process can lead to significantly smaller and less massive discs which are much less prone to gravitational instability. Similarly in the case of binary star formation we find that magnetic fields, overall, suppress fragmentation.  However these effects are found to be largely driven by magnetic pressure. The relative importance of magnetic tension is dependent on the orientation of the field with respect to the rotation axis, but can, with the right orientation, lead to a dilution of the magnetic pressure-driven suppression of fragmentation. 
\end{abstract}



\section{Introduction}
\label{sec:intro}
 Star forming regions are routinely observed to contain magnetic fields of strengths sufficient to play a significant role in the star formation process, delaying and perhaps preventing collapse \citep{crutcheretal04,hc05}. Furthermore magnetic fields are the main candidate for producing the ubiquitous jets and outflows observed emanating from star forming cores. For this reason it is crucial to be able to include the effects of magnetic fields into numerical models of the star forming process. Furthermore the role which magnetic fields play in the currently favoured `dynamical picture' of star formation \citep{mk04} is not well understood and only a limited number of numerical studies have been performed.
 
  The degree to which magnetic fields can counteract the gravitational instability is determined, for an enclosed region of gas threaded by a magnetic field, by the ratio of the mass contained within the region to the magnetic flux passing through the surface. This is referred to as the \emph{mass-to-flux} ratio, which for a spherical cloud takes the form:
\begin{equation}
\frac{M}{\Phi} \equiv \frac{M}{4\pi R^{2} B_{0}}.
\label{eq:masstoflux}
\end{equation} 
where $M$ is the mass contained within the cloud volume, $\Phi$ is the magnetic flux threading the cloud surface at radius $R$ assuming a uniform magnetic field $B_{0}$. The critical value of $M/\Phi$ below which a cloud will be supported against gravitational collapse is given by \citep[e.g.][]{ms76,mestel99,mk04}.
\begin{equation}
\left(\frac{M}{\Phi}\right)_{\rm crit} = \frac{2 c_{1}}{3} \sqrt{\frac{5}{\pi G \mu_{0} }},
\label{eq:mphicrit}
\end{equation}
where $G$ and $\mu_{0}$ are the gravitational constant and the permeability of free space respectively and $c_{1}$ is a constant determined numerically by \citet{ms76} to be $c_{1}\approx 0.53$. 
Star forming cores with mass-to-flux ratios less than unity are stable against collapse (``subcritical'') and conversely, cores with mass-to-flux ratios greater than unity (termed ``supercritical'') will collapse on the free-fall timescale.

 Magnetic fields also play a role in the transport of angular momentum away from star forming cores, both by the production of jets and outflows and also by `magnetic braking' -- that is regions of gas undergoing collapse and which therefore begin to rotate rapidly remain connected to more slowly rotating regions of gas by magnetic field lines, the induced tension of which acts to `brake' the star forming core. However, understanding the role of magnetic fields in the star formation process ultimately requires three dimensional, self-gravitating, magnetohydrodynamics (MHD) simulations. 

\section{Numerical method}
\label{sec:1}
 One of the most widely used methods for simulations of star formation is that of Smoothed Particle Hydrodynamics \citep[SPH -- for recent reviews see][]{Monaghan05,price04}, for the reason that the resolution automatically adapts to the mass distribution which is precisely where it is required in star formation simulations. The basis of the method is that fluid quantities are discretised onto a set of moving points (the `particles') which follow the fluid motion. For example, the density is computed as a sum over neighbouring particles in the form
\begin{equation}
\rho({\bf r}) = \sum_{j} m_{j} W(\vert {\bf r} - {\bf r}_{j}\vert, h), \label{eq:dens}
\end{equation}
where $m_{j}$ are the masses of neighbouring particles and $W$ is a weight function (the `smoothing kernel') -- something like a Gaussian although in practice a function which goes to zero at a finite radius (usually $2h$, where $h$ is the so called `smoothing length') is used for efficiency. It is a remarkable fact that, writing down the density in the form (\ref{eq:dens}) actually defines (almost) the \emph{entire numerical method}. What we mean by this is that, using only the density sum, it is possible to then self-consistently derive the equations of hydrodynamics in their numerical form with only the additional assumption of the first law of thermodynamics. This is possible because SPH can be derived from a Hamiltonian variational principle, where, for hydrodynamics, the Lagrangian takes the form
\begin{equation}
L = \sum_{j} m_{j} \left[ \frac12 v_{j}^{2}  - u_{j}(\rho_{j},s_{j}) \right],
\end{equation}
which is nothing more than the difference between the kinetic and potential (thermal) energies expressed as a sum over particles ($u$ refers to the thermal energy per unit mass which is assumed to be a function of the density and entropy). The Lagrangian can be written as a function of the particle co-ordinates using the density summation (\ref{eq:dens}) and the equations of hydrodynamics thus derived using the Euler-Lagrange equations. Simultaneous (unlike in a grid-based code) conservation of \emph{all} physical quantities (momentum, angular momentum, energy, entropy and even circulation -- see \citealt{mp01}) follows, reflecting the symmetries present in the Lagrangian.

This is a very powerful principle for development of SPH algorithms, as it means, fundamentally, that only one of two things can be changed (without losing some of the advantage of having a Hamiltonian method): either the density summation or the Lagrangian. An example of the former is the recent development of a self-consistent formulation in the presence of a spatially variable smoothing length due to \citet{sh02} and \citet{monaghan02} (see also \citet{pm07} for the extension of this formulation to gravitational force softening). Additional physics is introduced by changing the Lagrangian. 

 A method for magnetic fields in SPH can thus be derived using the Lagrangian
\begin{equation}
L = \sum_{j} m_{j} \left[ \frac12 v_{j}^{2}  - u_{j}(\rho_{j},s_{j}) - \frac12 \frac{B_{j}^{2}}{\mu_{0}\rho_{j}} \right],
\end{equation}
where the additional term is the magnetic energy. Such a derivation is presented by \citet{pm04b}. However, life is never that simple, and the derivation of a workable algorithm from that point is complicated by several factors. The first is that the momentum-conserving form of the SPMHD (Smoothed Particle Magnetohydrodynamics) equations proves to be (violently) unstable in the regime where the magnetic pressure exceeds the gas pressure. Second is that the Lagrangian says nothing about dissipation, which is a necessary introduction in order to resolve discontinuities in the flow (ie. shocks) which are made more complicated in MHD by the three different wave types (slow, Alfven and fast) and correspondingly complicated shock structures. The third complication is the use of a spatially variable smoothing length, although this can be incorporated into the Lagrangian derivation (and was done by \citealt{pm04b}). The fourth complication is that nasty fourth Maxwell equation, $\nabla\cdot {\bf B} = 0$, expressing the physical condition that no magnetic monopoles should exist. A lengthy description of methods for divergence cleaning in SPH which, for the most part, \emph{don't} work very well is given in \citet{pm05}. 

\begin{figure*}
\begin{center}
\includegraphics[width=0.8\textwidth]{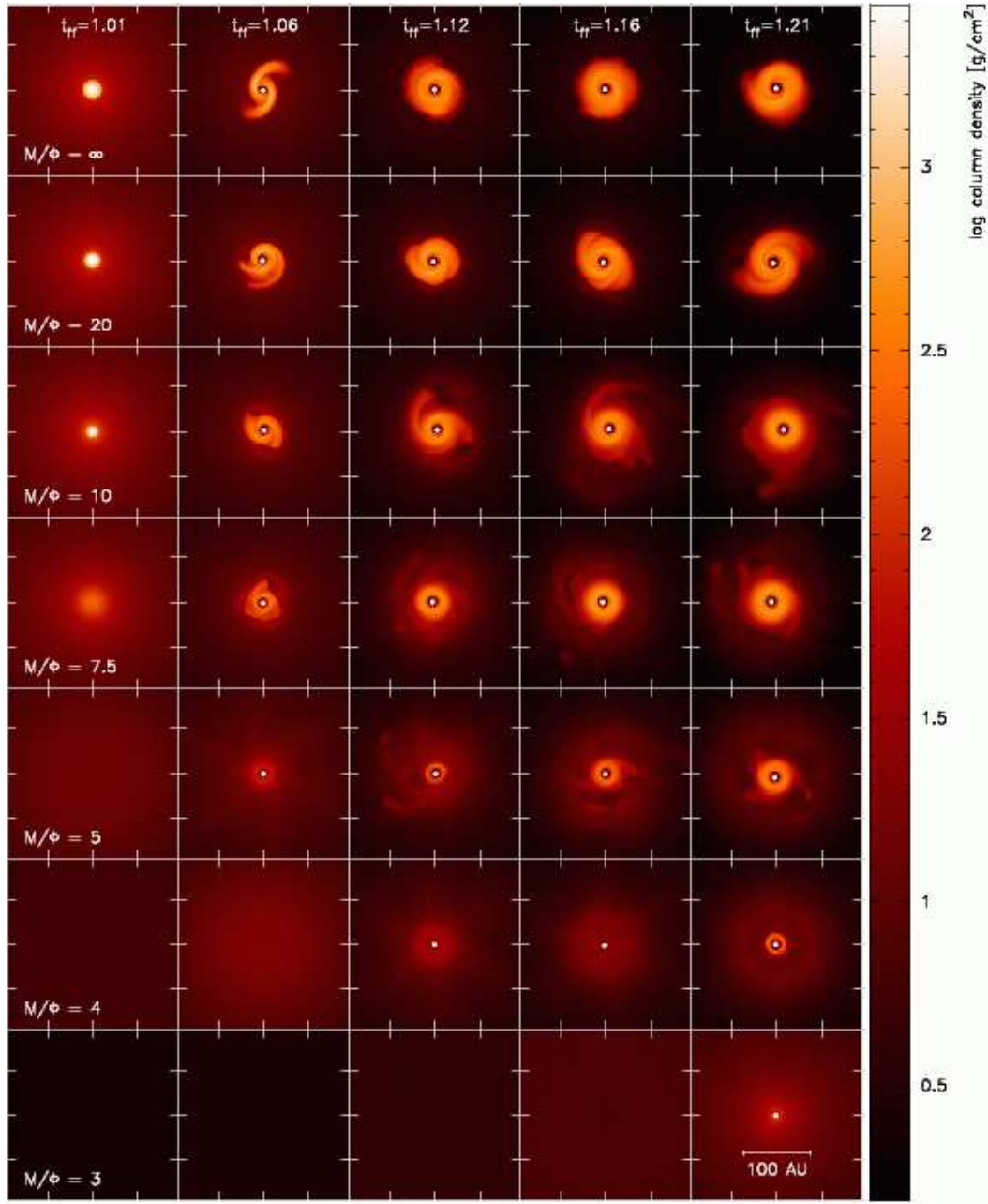}
\end{center}
\caption{The influence of magnetic fields on circumstellar disc formation. Plots show results of the single star collapse calculations at various times in the evolution (left to right) and for a series of runs of increasing magnetic field strength (top to bottom) with a field initially aligned with the rotation axis. The magnetic field delays the collapse and leads to smaller, less massive discs which are less prone to gravitational instability.}
\label{fig:singlestar_Bz}
\end{figure*}

 Overcoming each of these complications has been a somewhat long and tortuous process, consuming a number of otherwise-healthy PhD students about once per decade (of which I [Price] am the latest). Without boring the reader it is sufficient to say that (at least to our partial satisfaction) each of these issues has now been resolved. The resolutions are essentially 1) compromising the momentum-conserving force slightly in order to attain partial momentum-conservation but stability; 2) formulating dissipation terms for MHD following \citet{monaghan97} \citep[see][]{pm04a}; 3) deriving the variable smoothing length formulation from a Lagrangian and 4) using prevention not cure by formulating the magnetic field in a divergence free form using the `Euler potentials' $\alpha_{E}$ and $\beta_{E}$ such that ${\bf B} = \nabla \alpha_{E} \times \nabla\beta_{E}$. The latter has the further advantage that the Lagrangian evolution of these potentials for ideal MHD is zero, corresponding to the advection of magnetic field lines \citep{stern70}, although there are also disadvantages to their use. In practise we add artificial dissipation terms to the Euler potentials' evolution in order to resolve (and dissipate) strong gradients in the magnetic field (see \citealt{pb07} and \citealt{rp07} for more details of the Euler potentials formulation in SPH).
 
  The resulting method gives excellent results on a wide range of test problems used to benchmark recent grid-based MHD codes (see e.g. \citealt{price04,pm05,rp07}) and is here applied to star formation problems for the first time.

\section{Single and binary star formation}
 We consider the problem of single and binary star formation starting with a uniform density, spherical core ($R = 4\times 10^{16}$cm $ = 2674$AU, $M = 1M_{\odot}$) in solid body rotation and embedded in a warm, low density medium. The simulations use a barytropic equation of state which is isothermal ($T \sim 10K$) to a density of $\rho = 10^{-14}$g cm$^{-3}$ above which $\gamma = 7/5$ is assumed to approximately represent the transition to gas which is optically thick to radiation. We model the core using 300,000 SPH particles which is an order of magnitude more than is necessary to resolve the Jeans mass (and thus any fragmentation which occurs) in this problem \citep{bateburkert97}. 

 The important parameters to this problem are: i) the ratio of thermal to gravitational energy, $\alpha$ (expressing the competition between gravity and gas pressure), ii) the ratio of rotational to gravitational energy, denoted $\beta_{r}$ (gravity vs. rotation) and iii) the mass-to-flux ratio defined by Eqn. (\ref{eq:masstoflux}) (gravity vs. magnetic fields). We fix the first two and vary the latter. For the single star collapse calculations we consider the case $\alpha = 0.35$, $\beta_{r} = 0.005$ (given by $\Omega = 1.77\times 10^{-13}$ rad/s corresponding to a relatively slow rotation). 
 
 Using a supercritical magnetic field initially aligned with the rotation axis (that is, in the computational $z-$direction), at low field strengths we find that the field lines are dragged inwards by the collapse, whereas at high field strengths the collapse is directed along the magnetic field lines. Figure~\ref{fig:singlestar_Bz} shows the results of a series of simulations of increasing magnetic field strength (from top to bottom, where $M/\Phi$ refers to the mass-to-flux ratio in units of the critical value), shown at various times in units of the initial free-fall time (left to right). The simulations reveal a clear influence of the magnetic field on the formation of the circumstellar disc around the star forming core. In the hydrodynamics simulation (top row) the disc is very massive and as a result exhibits strong gravitational instabilities in the form of spiral arms.  For runs with increasing magnetic field strength the disc forms progressively later (e.g. no significant disc is visible up to $t_{ff} = 1.21$ in the $M/\Phi = 4$ run), and are substantially less massive, smaller and thus less prone to gravitational instability. The accretion rate of material onto the disc itself is also slower. This weakening of gravitational instabilities in discs by magnetic fields may have implications for the relative importance of gravitational instability as a planet formation mechanism. 

\begin{figure*}
\begin{center}
\includegraphics[width=0.8\textwidth]{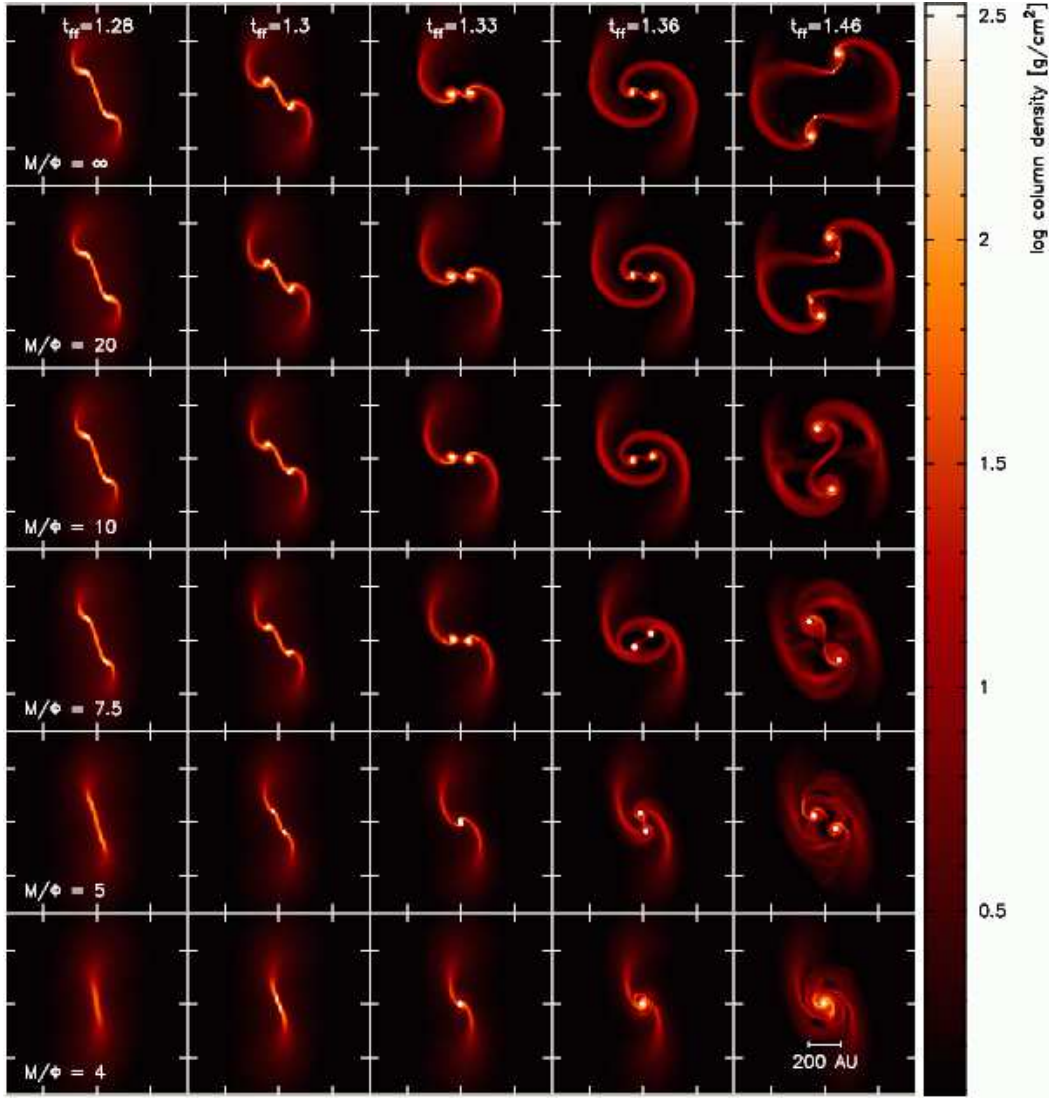}
\end{center}
\caption{Results of the binary collapse calculations at various times in the evolution (left to right) and for a series of runs of increasing magnetic field strength (top to bottom) with a field initially aligned with the rotation axis. The magnetic field delays the collapse and suppresses fragmentation in this case.}
\label{fig:binary_Bz}
\end{figure*}

 A similar trend is observed in binary star formation simulations (Figure~\ref{fig:binary_Bz}, which shows the results of simulations where an initial density perturbation of the form
\begin{equation}
\rho = \rho_{0}[1 + 0.1\cos{(2\phi)}],
\end{equation}
has been applied with $\alpha=0.26$ and $\beta_{r}=0.16$. Collapse is again delayed by the magnetic field and fragmentation is suppressed (that is, with increasing magnetic field the formation of a binary is suppressed and only a single star is formed). 
 
\begin{figure}[!ht]
\begin{center}
\includegraphics[width=\columnwidth]{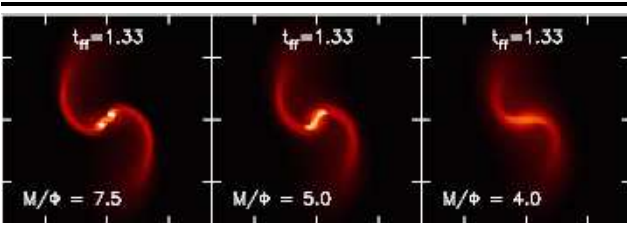}
\end{center}
\caption{Simulations of binary star formation performed with magnetic tension forces turned off (plots correspond to the central and bottom three panels of Figure~\ref{fig:binary_Bz}). The results are almost identical, indicating that it is magnetic pressure that is playing the dominant role in suppressing fragmentation. }
\label{fig:binary_pressureonly}
\end{figure}

  Whilst it is tempting to attribute these effects to the transport of angular momentum via magnetic braking (thus removing material from the central regions, and in the binary case removing angular momentum from the binary system), some further investigation is warranted. Figure~\ref{fig:binary_pressureonly} shows three of the binary collapse calculations at $t_{ff}=1.33$ where we have turned off the magnetic tension force (that is, using only a magnetic pressure term). The results are nearly (although not exactly) identical to those shown in Figure~\ref{fig:binary_Bz}, demonstrating that it is in fact magnetic pressure that is playing the dominant role in suppressing binary/disc formation by increasing the effective thermal energy (ie. the sum of both thermal and magnetic pressure) of the cloud. 
  
 A deeper investigation \citep{pb07} reveals that this result is slightly qualified by the initial orientation of the magnetic field with respect to the rotation axis. Simulations using a magnetic field initially aligned perpendicular to the rotation axis (ie. where the field lies initially in the orbital plane of the binary) show a much stronger contribution from magnetic tension which can in fact aid fragmentation (or rather, dilute the effect of magnetic pressure in suppressing fragmentation), confirming a scenario which had been suggested by Alan Boss \citep[e.g.][]{boss02} based on `approximate' MHD simulations. This occurs because, when the field is aligned perpendicular to the rotation axis it can form a ``magnetic cushion'' between overdense regions the tension force perpendicular to which prevents them from merging. 

 We are currently applying the method to simulations of star cluster formation from turbulent initial conditions \citep[as in][]{bbb03} which include the effects of magnetic fields.

\section{Summary}
 In summary, we have performed simulations of single and binary star formation using a recently developed method for Smoothed Particle Magnetohydrodynamics. We find that stronger magnetic fields result in a slower collapse, and that the extra support provided by magnetic pressure acts to suppress fragmentation and disc formation, resulting in smaller, less massive discs. The net result of this is that the presence of magnetic fields in the disc formation process can weaken gravitational instabilities in young, massive circumstellar discs which may have implications for the relative importance of gravitational instability as a planet formation mechanism.

\begin{acknowledgements}
DJP is supported by a UK PPARC postdoctoral research fellowship but would love to get a job in Australia. Calculations were performed using the School of Physics iMac cluster at the University of Exeter and the UK Astrophysical Fluids Facility (UKAFF). MRB is grateful for the support of a 
Philip Leverhulme Prize and a EURYI Award. 
\end{acknowledgements}

% BibTeX users please use the following style file
\bibliographystyle{Spr-mp-nameyear}

\bibliography{sph,mhd,starformation}

\begin{thebibliography}{}
\ifx \bisbn   \undefined \def \bisbn  #1{ISBN #1}   \fi
\ifx \binits  \undefined \def \binits#1{#1} \fi
\ifx \bauthor  \undefined \def \bauthor#1{#1} \fi
\ifx \batitle  \undefined \def \batitle#1{#1} \fi
\ifx \bjtitle  \undefined \def \bjtitle#1{#1} \fi
\ifx \bvolume  \undefined \def \bvolume#1{#1} \fi
\ifx \byear  \undefined \def \byear#1{#1} \fi
\ifx \bissue  \undefined \def \bissue#1{#1} \fi
\ifx \bfpage  \undefined \def \bfpage#1{#1} \fi
\ifx \blpage  \undefined \def \blpage #1{#1} \fi
\ifx \burl  \undefined \def \burl#1{#1} \fi
\ifx \binterref  \undefined \def \binterref#1{#1} \fi
\ifx \betal  \undefined \def \betal#1{#1} \fi
\ifx \binstitute  \undefined \def \binstitute#1{#1} \fi
\ifx \bctitle  \undefined \def \bctitle#1{#1} \fi
\ifx \beditor  \undefined \def \beditor#1{#1} \fi
\ifx \bpublisher  \undefined \def \bpublisher#1{#1} \fi
\ifx \bbtitle  \undefined \def \bbtitle#1{#1} \fi
\ifx \bedition  \undefined \def \bedition#1{#1} \fi
\ifx \bseriesno  \undefined \def \bseriesno#1{#1} \fi
\ifx \blocation  \undefined \def \blocation#1{#1} \fi
\ifx \bsertitle  \undefined \def \bsertitle#1{#1} \fi
\ifx \bsnm \undefined \def \bsnm#1{#1} \fi
\ifx \bsuffix \undefined \def \bsuffix#1{#1} \fi
\ifx \bparticle \undefined \def \bparticle#1{#1} \fi
\ifx \barticle \undefined \def \barticle#1{#1} \fi
\ifx \botherref \undefined \def \botherref #1{#1} \fi
\ifx \url \undefined \def \url#1{#1} \fi
\ifx \bchapter \undefined \def \bchapter#1{#1} \fi
\ifx \bbook \undefined \def \bbook#1{#1} \fi
\ifx \bcomment \undefined \def \bcomment#1{#1} \fi
\ifx \protect\citeauthoryear \undefined \def \protect\citeauthoryear#1{#1} \fi
\ifx \oauthor \undefined \def \oauthor#1{#1} \fi
\def \endbibitem {}

\bibitem[\protect\citeauthoryear{{Crutcher}
  \textit{et~al.}}{2004}]{crutcheretal04}
\begin{barticle}
\bauthor{\bsnm{{Crutcher}},~\binits{R.M.}},
  \bauthor{\bsnm{{Nutter}},~\binits{D.J.}},
  \bauthor{\bsnm{{Ward-Thompson}},~\binits{D.}},
  \bauthor{\bsnm{{Kirk}},~\binits{J.M.}}:
\batitle{{SCUBA Polarization Measurements of the Magnetic Field Strengths in
  the L183, L1544, and L43 Prestellar Cores}}.
\bjtitle{ApJ} \bvolume{600},  \bfpage{279}--\blpage{285} (\byear{2004}).
\end{barticle}
\endbibitem

\bibitem[\protect\citeauthoryear{{Heiles} and {Crutcher}}{2005}]{hc05}
\begin{botherref}
\oauthor{\bsnm{{Heiles}}~\binits{C}}, \oauthor{\bsnm{{Crutcher}}~\binits{R}}:
{Magnetic Fields in Diffuse HI and Molecular Clouds}.
In: {Wielebinski}, R., {Beck}, R. (eds.) LNP Vol. 664: Cosmic Magnetic Fields.
  p. 137. (2005)
\end{botherref}
\endbibitem

\bibitem[\protect\citeauthoryear{{Mac Low} and {Klessen}}{2004}]{mk04}
\begin{barticle}
\bauthor{\bsnm{{Mac Low}},~\binits{M.}},
  \bauthor{\bsnm{{Klessen}},~\binits{R.S.}}:
\batitle{{Control of star formation by supersonic turbulence}}.
\bjtitle{Rev. Mod. Phys.} \bvolume{76},  \bfpage{125}--\blpage{194}
  (\byear{2004})
\end{barticle}
\endbibitem

\bibitem[\protect\citeauthoryear{{Mouschovias} and {Spitzer}}{1976}]{ms76}
\begin{barticle}
\bauthor{\bsnm{{Mouschovias}},~\binits{T.C.}},
  \bauthor{\bsnm{{Spitzer}},~\binits{L.}\bsuffix{Jr.}}:
\batitle{{Note on the collapse of magnetic interstellar clouds}}.
\bjtitle{ApJ} \bvolume{210},  326 (\byear{1976})
\end{barticle}
\endbibitem

\bibitem[\protect\citeauthoryear{{Mestel}}{1999}]{mestel99}
\begin{botherref}
\oauthor{\bsnm{{Mestel}}~\binits{L}}:
{Stellar magnetism}. Oxford: Clarendon (1999)
\end{botherref}
\endbibitem

\bibitem[\protect\citeauthoryear{{Monaghan}}{2005}]{Monaghan05}
\begin{barticle}
\bauthor{\bsnm{{Monaghan}},~\binits{J.J.}}:
\batitle{Smoothed particle hydrodynamics}.
\bjtitle{Rep. Prog. Phys.} \bvolume{68}(\bissue{8}),
  \bfpage{1703}--\blpage{1759} (\byear{2005}).
\end{barticle}
\endbibitem

\bibitem[\protect\citeauthoryear{{Price}}{2004}]{price04}
\begin{botherref}
\oauthor{\bsnm{{Price}}~\binits{DJ}}:
{Magnetic fields in Astrophysics}.
PhD thesis, University of Cambridge, Cambridge, UK. astro-ph/0507472 (2004)
\end{botherref}
\endbibitem

\bibitem[\protect\citeauthoryear{{Monaghan} and {Price}}{2001}]{mp01}
\begin{barticle}
\bauthor{\bsnm{{Monaghan}},~\binits{J.J.}},
  \bauthor{\bsnm{{Price}},~\binits{D.J.}}:
\batitle{{Variational principles for relativistic smoothed particle
  hydrodynamics}}.
\bjtitle{MNRAS} \bvolume{328},  \bfpage{381}--\blpage{392} (\byear{2001})
\end{barticle}
\endbibitem

\bibitem[\protect\citeauthoryear{{Springel} and {Hernquist}}{2002}]{sh02}
\begin{barticle}
\bauthor{\bsnm{{Springel}},~\binits{V.}},
  \bauthor{\bsnm{{Hernquist}},~\binits{L.}}:
\batitle{{Cosmological smoothed particle hydrodynamics simulations: the entropy
  equation}}.
\bjtitle{MNRAS} \bvolume{333},  \bfpage{649}--\blpage{664} (\byear{2002})
\end{barticle}
\endbibitem

\bibitem[\protect\citeauthoryear{{Monaghan}}{2002}]{monaghan02}
\begin{barticle}
\bauthor{\bsnm{{Monaghan}},~\binits{J.J.}}:
\batitle{{SPH compressible turbulence}}.
\bjtitle{MNRAS} \bvolume{335},  \bfpage{843}--\blpage{852} (\byear{2002})
\end{barticle}
\endbibitem

\bibitem[\protect\citeauthoryear{{Price} and {Monaghan}}{2007}]{pm07}
\begin{barticle}
\bauthor{\bsnm{{Price}},~\binits{D.J.}},
  \bauthor{\bsnm{{Monaghan}},~\binits{J.J.}}:
\batitle{{An energy-conserving formalism for adaptive gravitational force
  softening in smoothed particle hydrodynamics and N-body codes}}.
\bjtitle{MNRAS} \bvolume{374},  \bfpage{1347}--\blpage{1358} (\byear{2007}).
\end{barticle}
\endbibitem

\bibitem[\protect\citeauthoryear{{Price} and {Monaghan}}{2004}]{pm04b}
\begin{barticle}
\bauthor{\bsnm{{Price}},~\binits{D.J.}},
  \bauthor{\bsnm{{Monaghan}},~\binits{J.J.}}:
\batitle{{Smoothed Particle Magnetohydrodynamics II. Variational principles and
  variable smoothing length terms}}.
\bjtitle{MNRAS} \bvolume{348},  139 (\byear{2004})
\end{barticle}
\endbibitem

\bibitem[\protect\citeauthoryear{{Price} and {Monaghan}}{2005}]{pm05}
\begin{barticle}
\bauthor{\bsnm{{Price}},~\binits{D.J.}},
  \bauthor{\bsnm{{Monaghan}},~\binits{J.J.}}:
\batitle{{Smoothed Particle Magnetohydrodynamics III. Multidimensional tests
  and the $\nabla\cdot{\bf B}=0$ constraint}}.
\bjtitle{MNRAS} \bvolume{364},  \bfpage{384}--\blpage{406} (\byear{2005})
\end{barticle}
\endbibitem

\bibitem[\protect\citeauthoryear{{Monaghan}}{1997}]{monaghan97}
\begin{barticle}
\bauthor{\bsnm{{Monaghan}},~\binits{J.J.}}:
\batitle{{SPH and Riemann Solvers}}.
\bjtitle{J. Comp. Phys.} \bvolume{136},  \bfpage{298}--\blpage{307}
  (\byear{1997})
\end{barticle}
\endbibitem

\bibitem[\protect\citeauthoryear{{Price} and {Monaghan}}{2004}]{pm04a}
\begin{barticle}
\bauthor{\bsnm{{Price}},~\binits{D.J.}},
  \bauthor{\bsnm{{Monaghan}},~\binits{J.J.}}:
\batitle{{Smoothed Particle Magnetohydrodynamics I. Algorithms and tests in one
  dimension}}.
\bjtitle{MNRAS} \bvolume{348},  \bfpage{123}--\blpage{138} (\byear{2004})
\end{barticle}
\endbibitem

\bibitem[\protect\citeauthoryear{{Stern}}{1970}]{stern70}
\begin{barticle}
\bauthor{\bsnm{{Stern}},~\binits{D.P.}}:
\batitle{{Euler Potentials}}.
\bjtitle{Am. J. Phys.} \bvolume{38},  \bfpage{494}--\blpage{501} (\byear{1970})
\end{barticle}
\endbibitem

\bibitem[\protect\citeauthoryear{{Price} and {Bate}}{2007}]{pb07}
\begin{barticle}
\bauthor{\bsnm{{Price}},~\binits{D.J.}},
  \bauthor{\bsnm{{Bate}},~\binits{M.R.}}:
\batitle{{The impact of magnetic fields on single and binary star formation}}.
\bjtitle{MNRAS} \bvolume{377},  \bfpage{77}--\blpage{90} (\byear{2007})
\end{barticle}
\endbibitem

\bibitem[\protect\citeauthoryear{{Rosswog} and {Price}}{2007}]{rp07}
\begin{botherref}
\oauthor{\bsnm{{Rosswog}}~\binits{S}}, \oauthor{\bsnm{{Price}}~\binits{DJ}}:
{MAGMA: A 3D, Lagrangian Magnetohydrodynamics code for merger applications}.
  MNRAS, accepted (2007, in press)
\end{botherref}
\endbibitem

\bibitem[\protect\citeauthoryear{{Bate} and {Burkert}}{1997}]{bateburkert97}
\begin{barticle}
\bauthor{\bsnm{{Bate}},~\binits{M.R.}},
  \bauthor{\bsnm{{Burkert}},~\binits{A.}}:
\batitle{{Resolution requirements for smoothed particle hydrodynamics
  calculations with self-gravity}}.
\bjtitle{MNRAS} \bvolume{288},  \bfpage{1060}--\blpage{1072} (\byear{1997})
\end{barticle}
\endbibitem

\bibitem[\protect\citeauthoryear{{Boss}}{2002}]{boss02}
\begin{barticle}
\bauthor{\bsnm{{Boss}},~\binits{A.P.}}:
\batitle{{Collapse and Fragmentation of Molecular Cloud Cores. VII. Magnetic
  Fields and Multiple Protostar Formation}}.
\bjtitle{ApJ} \bvolume{568},  \bfpage{743}--\blpage{753} (\byear{2002})
\end{barticle}
\endbibitem

\bibitem[\protect\citeauthoryear{{Bate}, {Bonnell}, and {Bromm}}{2003}]{bbb03}
\begin{barticle}
\bauthor{\bsnm{{Bate}},~\binits{M.R.}},
  \bauthor{\bsnm{{Bonnell}},~\binits{I.A.}},
  \bauthor{\bsnm{{Bromm}},~\binits{V.}}:
\batitle{{The formation of a star cluster: predicting the properties of stars
  and brown dwarfs}}.
\bjtitle{MNRAS} \bvolume{339},  \bfpage{577}--\blpage{599} (\byear{2003})
\end{barticle}
\endbibitem

\end{thebibliography}

\end{document}